\documentstyle[epsfig,12pt]{article}
\begin{document}
\title{ Effect of Internal Breeding of Tritium and Helium-3 on the Ignition of an ICF Fuel Pellet }
\author{T. Koohrokhi$^{1,}$\thanks{email:{\tt T.Koohrokhi@gu.ac.ir}} and R. Azadifar$^{2,}$\thanks{email:{\tt R.Azadifarr@gmail.com}} }
\maketitle \centerline{\it $^{1,2}$Faculty of Sciences, Golestan
University, Shahid Beheshti Street, P.O. Box 155, Gorgan, IRAN}
 \vspace{15pt}
\newpage

%
%
\begin{abstract}
Self-heating condition and following ignition in an Inertial
Confinement Fusion (ICF) fuel pellet is evaluated by calculating the
power equations, dynamically. In fact, the self-heating condition is
a criterion that determines the minimum parameters of a fuel (such
as temperature, density and areal density) that can be ignited.
Deuterium is the main component of ICF fuels as large amounts of it
are naturally available. In addition, the use of deuterium as a fuel
in ICF causes the production of tritium and helium-3. However, pure
deuterium has a high ignition temperature (T$\geq$40 keV) which
makes it inefficient.
 In this paper, the power equations are solved, dynamically, and
 it has been indicated
 that internal tritium and helium-3 production at early evolution of compressed deuterium fuel
  causes ignition at lower predicted temperatures.\\

\end{abstract}

%
%
\section{Introduction}
Inertial Fusion Energy (IFE) research is approaching a critical
juncture in its history with high expectations that ignition and
physical feasibility will be demonstrated soon in related facilities
around the world$^{1,2}$. The main components of the ICF fuel
pellets are the hydrogen isotopes, Deuterium and Tritium (DT). The
fusion reaction of deuterium and tritium turns out to be the easiest
approach to fusion because of a relatively large cross section and a
very high mass defect$^3$.
 As a result the DT mixture is the fuel with the lowest ignition
temperature and has the highest specific yield. Since the most
practical nuclear fusion reaction for power generation seems to be
the DT reaction, the sources of these fuels are important. The
deuterium part of the fuel does not pose a great problem because
about 1 part in 5000 of the hydrogen in seawater is deuterium$^{4}$.

While large amounts of deuterium are naturally available in ocean
water and then makes up 156 ppm of hydrogen on earth, tritium can be
bred from the lithium isotope $^{6}\textrm{Li}$ which is also
available in great quantities on earth$^{5}$. However, the tritium
is an unstable artificial isotope, decaying to $^{3}\textrm{He}$
with a half life of 12.3 years, and as a result needs to be produced
within the DT fuel cycle$^{6}$. Therefore, the tritium production
has most significant radiological problem in future DT fusion
reactors$^7$. In a more mature fusion power economy, the tritium
breeding ratio (TBR) will be adjusted to a point closer to unity
with just enough extra to cover decay and any losses$^8$.
Furthermore, energetic neutron (14.1 MeV) that it yield from DT
reaction is another problem.

Helium-3 is a helium isotope that is light and non-radioactive.
Nuclear fusion reactors using helium-3 could provide a highly
efficient form of nuclear power with virtually no waste and
negligible radiation. Nuclear fusion using helium-3 would be
cleaner, as it does not produce any spare neutrons. Although the
helium-3 is almost non-existent on earth, it does exist on the
moon$^{9}$. Lacking an atmosphere, the moon has been bombarded for
billions of years by solar winds carrying helium-3$^{10}$.

One requirement of traditional inertial fusion energy (IFE) power
plant designs is the need to breed large quantities of tritium to
replace that which is burned. By using the pure deuterium as fuel,
the complex tritium breeding blanket would not be necessary and the
tritium inventory in the reactor would be substantially reduced.
This means increased safety and smaller environmental impact in case
of accidents. In addition, the reduced number of 14.1 MeV neutrons
and the softer neutron spectrum would ease the problems related to
neutron induced damage, mostly due to neutrons with energy above 4-5
MeV. However, burning of pure deuterium, requiring very high
temperature and large areal density $\rho R$, and the ignition of
pure deuterium seems unrealistic in ICF because of high ignition
temperature (T$\geq$40 keV)$^{11,12}$. It has been shown that
laser-induced nuclear fusion processes can be occur in ultra-dense
deuterium$^{13-15}$. The deuteron fast ignition is an alternative
way to nuclear energy production by deuterium fusion$^{16-18}$. In
this paper, we indicate that the deuterium fuel can be ignited below
the deuterium ignition temperature via a process which will be
explained through out the paper.


\section{Self-Heating Following Ignition}
In laser-driven inertial confinement fusion, spherical capsules are
compressed and heated to high enough temperatures and densities for
fusion reactions to occur$^{11,12}$. First, the laser irradiation
leads to surface ablation and drives the fuel implosion. As the
imploding material stagnates in the center, its kinetic energy is
converted into internal energy. At this time, the fuel consists of a
highly compressed shell enclosing a hot spot of igniting fuel in the
center (Fig. 1). A burn wave starting from the hot spot then ignites
the whole fuel, which explodes. Since ignition is occurred from
center of the hot spot, this scheme is named "central hot spot".

There are two forms of the imploded fuel at the ignition conditions,
namely isobaric and isochoric. The isochoric assembly is a
configuration in which hot spot and cold fuel have the same uniform
density. This is in contrast to the isobaric hot spot scenario,
where it is essential that the hot spot area and the surrounding
cold fuel remain in pressure equilibrium during compression. In
fact, the pressure distribution over the stagnating fuel is not
really uniform, but drops to low values at the outer boundary. The
isobaric model is therefore overestimating the energy invested into
the cold fuel layer. Simulations of the ignition conditions in
($\rho$R, T )-space show that the isochoric compression requires a
higher $\rho$R than in the isobaric case, because the cold fuel
layer works as a tamper suppressing the expansion of the central
plasma regions$^{19}$. Nevertheless, isochoric conditions convenient
for the alternative fast ignition approach that it can lead to a
fuel gain a factor 2-3 larger than the usual hydrodynamic ignition
by a central isobaric hot spot$^{20}$.


\subsection{Hot Spot Self-Heating Condition}
Let us consider the schematic case illustrated in Fig. 1. The laser
pulse has compressed fuel homogeneously to density $\rho=5000$, and
now the thermonuclear reaction starts in a hot spot with a areal
density $\rho R$ and temperature T. The rate of change of the
internal energy density
$E$ of the hot spot is,\\
\begin{equation}
    \frac{dE}{dt}=P_{f}-P_{B}-P_{C}-P_{e}-P_{m}
\end{equation}
$P_{f}$ is the power density deposited by the fusion products,
$P_{m}$ is the contribution due to mechanical work, $P_{C}$ is the
power density due to inverse Compton scattering, $P_{B}$ and $P_{e}$
are, respectively, the power densities lost by radiation and by
thermal conduction. Each power has been introduced in its details in
previous studies$^{21,22}$. According to Eq. (1), the hot spot
temperature increases when,\\
\begin{equation}
    P_{f}>P_{B}+P_{C}+P_{e}+P_{m}
\end{equation}
that is, the power deposition by fusion products exceeds the sum of
all power losses. This equation is solved numerically for pure
deuterium fuel with $\rho=5000~(\textrm{gcm}^{-3})$ density. The
dashed area in Fig. 2 displays the region in the $\rho R-T$ plane,
where Eq. (2) is satisfied and solid line represents a solution when
the inequality is replaced by an equality. Instantaneous power
balance allows to determine whether a hot spot cools or heats and
thus, the Eq. (2) is called hot spot self-heating condition. Below
the self-heating area power losses dominate the power balance.
Indeed, if the initial parameters lie within this region or are even
just on the boundary then expect the temperature increases and thus
allows more fusion reactions to take place. Note that this equation
is a static equation and temperature of electrons and ions in its
solution are considered identical.

\begin{figure}[htp]
\begin{center}\includegraphics{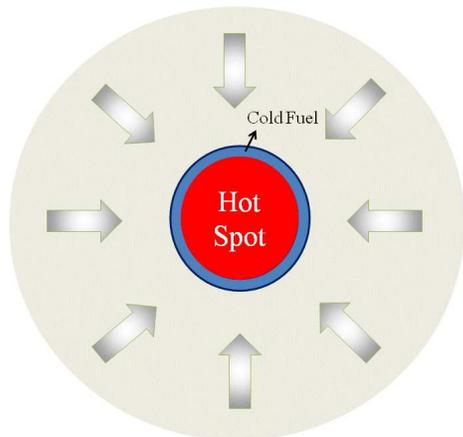} \vspace{6.5cm}
\end{center}
 \caption{\small {Schematic picture of the compressed ICF fuel pellet to form the hot spot sourended by cold fuel for ignition.   }}
\end{figure}

\begin{figure}[htp]
\begin{center}\includegraphics{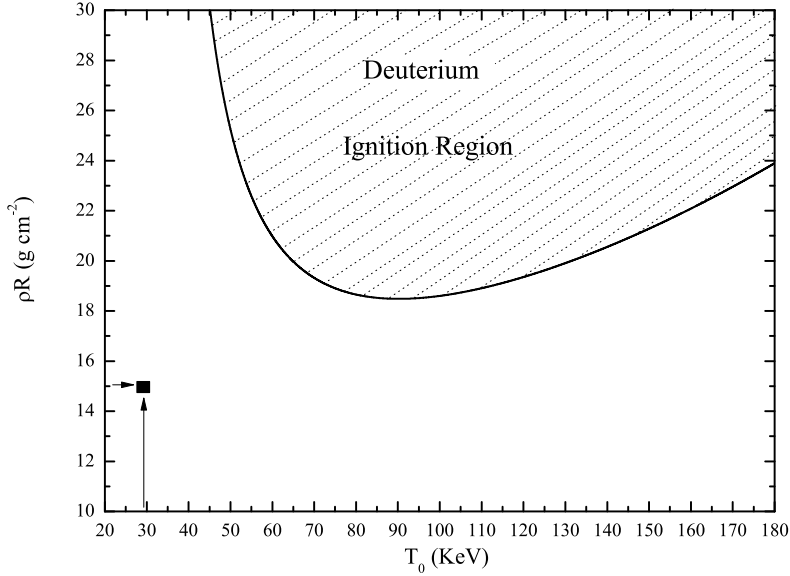} \vspace{6.5cm}
\end{center}
 \caption{\small {Self-heating conditions,
Eq. (2), in the $\rho$ R-T plane for a deuterium hot spot in
isochoric configuration at $\rho=5000~(\textrm{gcm}^{-3})$ density.
}}
\end{figure}


\subsection{Ignition and Time Evolution}
Now we perform a time-dependent calculation of the fusion processes
of a $\textrm{DT}_{x}~^{3}\textrm{He}_{y}$ configuration fuel
pellet, where x is the ratio of the tritium to deuterium and y is
the ratio of the helium-3 to deuterium particle numbers at initial
time (t = 0). When the fusion reactions taking place, the total
number density of particles of
species $k$, $n_{k}$, is governed by the equation,\\
\begin{equation}
      \frac{dn_{k}}{dt}=\sum_{j} a_{k}^{j}n_{j(1)}n_{j(2)}<\sigma v >_{j}
\end{equation}
where $<\sigma v>_{j}$ is the Maxwell averaged reaction rate of
reaction $j$, and $a_{k}^{j}$ is the number of particles of species
$k$ created or destroyed in the reaction $j$. Six species are
considered in this calculation: D, $^{3}\textrm{He}$, T, p,
$^{4}\textrm{He}$ and n.

For the ignition and time evolution, it is important to take into
account the different evolutions of the ion and electron
temperatures. The electron energy losses by the bremsstrahlung
radiation, inverse Compton effect and thermal conduction. Therefore,
the ion temperature will be higher than the electron temperature and
an energy flow from the ions to the electrons by collisions is
expected. The corresponding energy balance involving the ion and the
electron internal energy densities $E_{i}$ and $E_{e}$,
respectively, are described by the following equations,\\
\begin{equation}
      \frac{dE_{i}}{dt}=\frac{3}{2}\frac{d}{dt}(n_{i}T_{i})=\sum_{j}\sum_{k} (1-\eta_{k}^{j})f_{k}^{j}P_{fk}-P_{ie}-P_{mi}
      \end{equation}
\begin{equation}
      \frac{dE_{e}}{dt}=\frac{3}{2}\frac{d}{dt}(n_{e}T_{e})=\sum_{j}\sum_{k}(1-\eta_{k}^{j})
      (1-f_{k}^{j})P_{fk}+P_{ie}-P_{B}-P_{C}-P_{e}-P_{me}
\end{equation}
where $T_{i}$ ($n_{i}$) and $T_{e}$ ($n_{e}$) are the ion and
electron temperature (number density), respectively, $\eta_{k}^{j}$
is the energy leakage probability of the product $k$ created in the
reaction $j$ and $f_{k}^{j}$ is the fraction of the energy of the
product $k$ created in the reaction $j$ that is deposited into the
plasma ions$^23$. Since the electron and ion temperatures are
different, ion-electron power density, $P_{ie}$, is inserted to
account for the flow of energy between them.

let us consider, for example, pure deuterium fuel with initial
configuration of density $\rho_{0}=$ 5000 $\textrm{gcm}^{-3}$, areal
density $\rho_{0} R_{0} =$ 15 $\textrm{gcm}^{-2}$ and initial ion,
electron temperatures given by $T_{i} = T_{e} =$ 29 keV.
Corresponding the Fig. 2, these parameters are out of self-heating
conditions (black point in Fig. 2) and thus the temperature expected
to decrease and the fuel can not achieve ignition. To evaluate the
fuel evolution, the fig. 3 is drawn by solving the equations 3, 4
and 5, numerically. These equations are coupled and must be solved
simultaneously. The figure shows that the temperature first
decreases slightly and then soon (t$\approx$1.5 ps) increases to
reach very high values. Therefore, ignition can also be achieved by
a hot spot with initial conditions outside the self-heating region.
In this case, at first the hot spot cools, and later self-heats and
ignites. It seems this early ignition at time evolution of
compressed fuel is a surprising event that need more exact analysis.

To find out what happen, the figures 4 and 5 are drawn until
ignition time ($0\leq t\leq1.5$~ ps). These figures show that
density is decreased because of fuel expansion (fig. 4), and the
tritium and helium-3 are produced duo to deuterium reactions
(D(D,p)T and D(D,n)~$^{3}\textrm{He}$), (fig. 5). In general, at the
ignition time the fuel parameters are,\\

\begin{displaymath}
\left\{ \begin{array}{ll}
    \mbox{$N_{D}= 1.65\times 10^{20} $}\\
    \mbox{$N_{T}= 4.14\times 10^{17} $}\\
    \mbox{$N_{^{3}He}= 7.50\times 10^{17} $}\\
    \mbox{$\rho=$ 4984 $\textrm{gcm}^{-3}$}\\
    \mbox{$\rho R =$ 14.9 $\textrm{gcm}^{-2}$}\\
    \mbox{$T_{i}=$ 25.35 keV}
                    \end{array}
                    \right.
\end{displaymath}

This means the pure deuterium fuel
($\textrm{DT}_{x=0}~^{3}\textrm{He}_{y=0}$) with density $\rho_{0}=$
5000 $\textrm{gcm}^{-3}$, areal density $\rho_{0} R_{0} =$ 15
$\textrm{gcm}^{-2}$ and temperature $T_{0}=$ 29 keV at the initial
time (t=0 ps), is converted to the fuel
$\textrm{DT}_{x=0.0025}~^{3}\textrm{He}_{y=0.0045}$ with density
$\rho=$ 4984 $\textrm{gcm}^{-3}$, areal density $\rho R =$ 14.9
$\textrm{gcm}^{-2}$ and temperature $T_{i}=$ 25.35 keV at ignition
time (t=1.5 ps). Now we recalculate the self-heating condition (Eq.
(2)) for this new configuration. Figure 6 is drawn for the time of
(t=1.5 ps) and shows that Eq. (2) is satisfied (black point) and
therefore the temperature can increase and fuel
will ignite.\\

\begin{figure}[htp]
\begin{center}\includegraphics{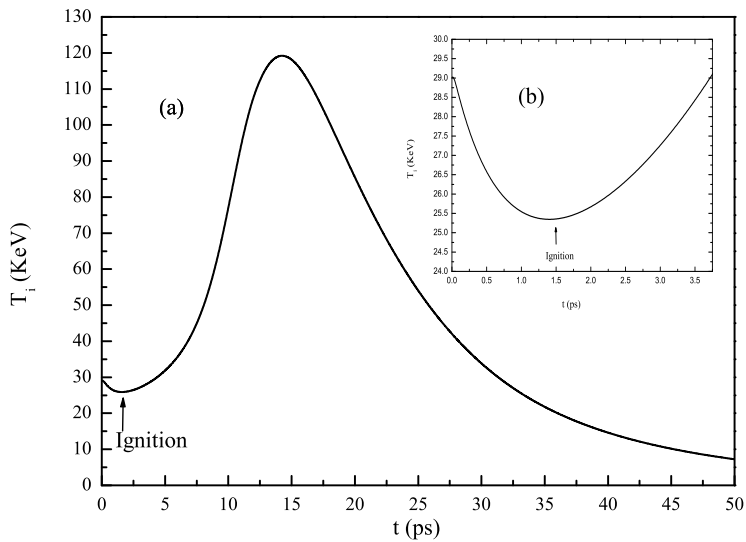} \vspace{7.5cm}
\end{center}
 \caption{\small {Ion temperature evolution of deuterium fuel
 as a function of burning time.
 The density of the
initial pellet is $\rho_{0}=$ 5000 $\textrm{gcm}^{-3}$. The areal
density is $\rho_{0}R_{0}=$ 15 $\textrm{gcm}^{-2}$, and the initial
temperature is $T_{0}=$29 keV. }}
\end{figure}

\begin{figure}[htp]
\begin{center}\includegraphics{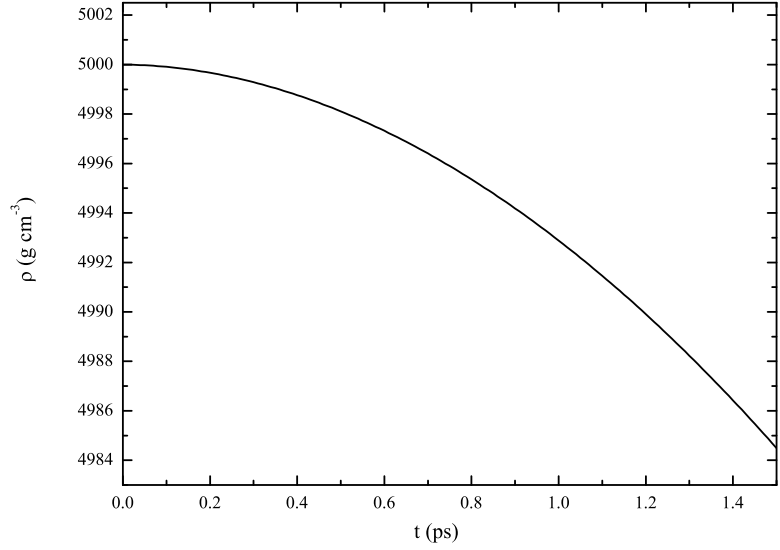} \vspace{7.5cm}
\end{center}
 \caption{\small {Density changes during 1.5 ps.   }}
\end{figure}

\begin{figure}[htp]
\begin{center}\includegraphics{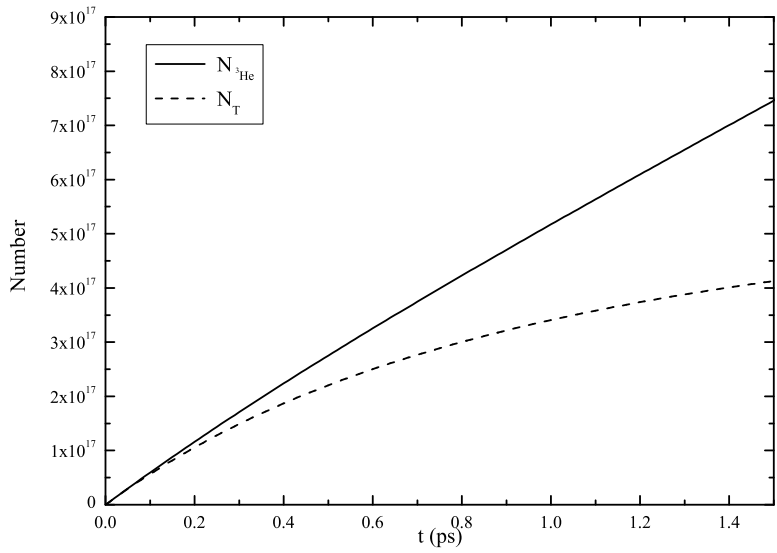} \vspace{7.5cm}
\end{center}
 \caption{\small {Contents of tritium and helium-3 during 1.5 ps.   }}
\end{figure}

\begin{figure}[htp]
\begin{center}\includegraphics{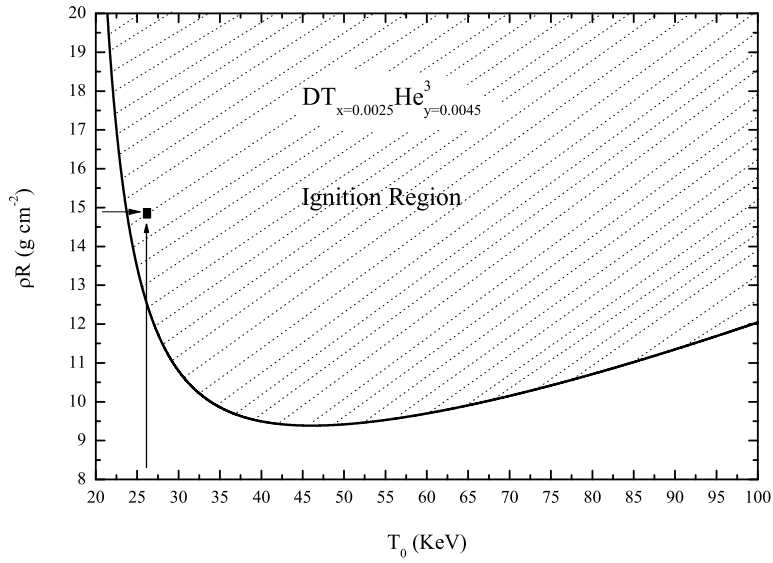} \vspace{7.5cm}
\end{center}
 \caption{\small {Self-heating conditions,
Eq (2), in the $\rho$ R-T plane for fuel
$\textrm{DT}_{x=0.0025}~^{3}\textrm{He}_{y=0.0045}$ with density
$\rho=$ 4984 $\textrm{gcm}^{-3}$, areal density $\rho R =$ 14.9
$\textrm{gcm}^{-2}$ and temperature $T_{i}=$ 25.35 keV for time
t=1.5 ps. }}
\end{figure}

\begin{figure}[htp]
\begin{center}\includegraphics{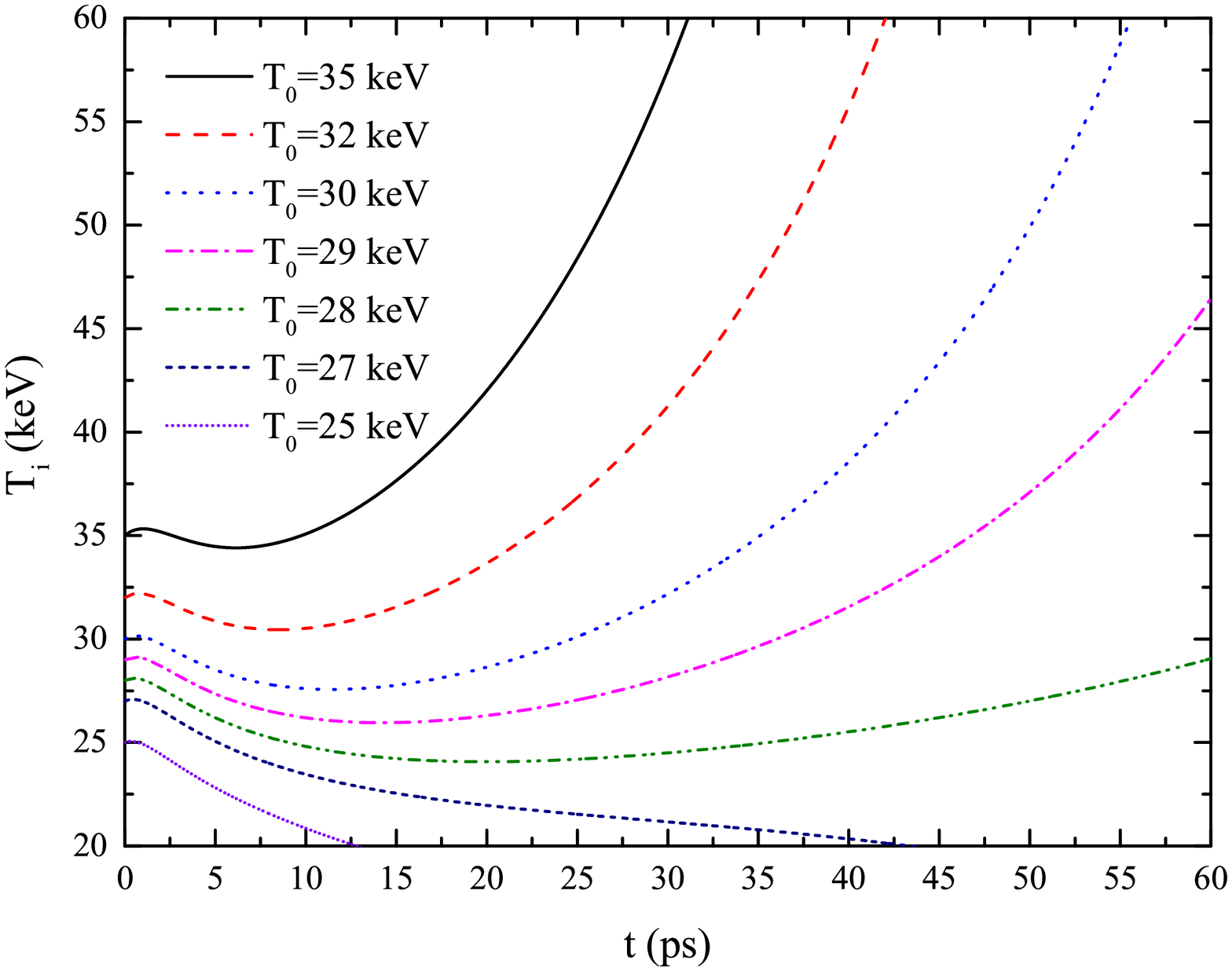} \vspace{7.5cm}
\end{center}
 \caption{\small {Ion temperature evolution of deuterium fuel
 as a function of burning time for different initial
temperature.
 The density and areal
density of initial fuel pellet is $\rho_{0}=$ 500
$\textrm{gcm}^{-3}$, $\rho_{0}R_{0}=$ 1.5 $\textrm{gcm}^{-2}$,
respectively. }}
\end{figure}


\section{Discussions}

Despite advantages of the use of deuterium such as stability and
natural availability, the pure deuterium fuel can not ignite easily.
A rough estimate of the energy required for ignition indicates that
the ignition of pure deuterium requires about $10^{4}$ times more
energy than ignition of DT, compressed to the same density. In this
paper, it have been mentioned that the ignition can be occurred even
blow the predicted temperature by self-heating condition.

The similar results may be found in previous works that can be
explained as follows: charged particles and electron conduction heat
a thin layer of the surrounding cold matter, which heats up and
ablates. The mass of the hot spot therefore increases in time, and
part of the energy lost by the hot spot is recovered. It may then
happen that a hot spot initially cools, while its mass and $\rho$R
increase. In such a way, the hot spot captures a larger fraction of
the charged particles, and may eventually heat up again and ignite
$^{24,25}$.

However, corresponding this paper, the early ignition at time
 evolution of the hot spot is occurred in a different manner. In fact the deuterium reactions
 (D(D,p)T and D(D,n)~$^{3}\textrm{He}$) produce tritium and helium-3 and these ions play
 a catalyzer$^{26}$ role via secondary reactions (T(D,n)~$^{4}\textrm{He}$
 and $^{3}\textrm{He}$(D,p)~$^{4}\textrm{He}$)$^{27}$. These reactions have large reaction rate that
 can raise the temperature fast enough to cause ignition. Since the reaction rate of T(D,n)~$^{4}\textrm{He}$ reaction is larger than
$^{3}\textrm{He}$(D,p)~$^{4}\textrm{He}$ reaction and then the final
content of tritium is smaller than helium-3, (fig. 5). However, pure
deuterium can be used as breeder fuel in concept of fusion plasma
that ignited by lower temperatures are predicted by the static
equations.

The initial density $\rho=5000~\textrm{gcm}^{-3}$ has been selected
in accordance with previous theoretical studies$^{27-30}$. It seems
this density is impossible to achieve by present technology. The
standard scheme uses densities up to 1000 $\textrm{gcm}^{-3}$ in the
(cold) main fuel while fast ignition should work with 300-500
$\textrm{gcm}^{-3}$ and $\rho R\leq 1.5~\textrm{gcm}^{-2}$. In the
Fig. 7, ion temperature evolution of deuterium fuel is shown as a
function of burning time for different initial temperatures and
density  $\rho=500~\textrm{gcm}^{-3}$. It can be seen that the fuel
pellet with $T_{0}\geq 28~\rm (keV)$ can be ignited with the same
behavior described above. By assuming the fuel plasma as a uniform
sphere with Maxwellian velocity distribution and initial conditions
as, $\rho=500~\textrm{gcm}^{-3}$, $T_{0}=28~\rm (keV)$ and $\rho
R=1.5~\textrm{gcm}^{-2}$, the internal energy $E_{i}=3/2NT\simeq
227~\rm (keV)$ should be provided by driver beams. These
calculations have been made as simple as possible to show the
breeding effect of pure deuterium fuel. More precise calculations
are needed containing hydrodynamic simulations. These more detailed calculations will be the subject of further studies by the author.\\


\section{Acknowledgments}
This work was supported by the Golestan University (Iran) under
contract of Research Project No. 92/71/20809.


\clearpage
\end{document}